\documentclass[twocolumn]{aastex631}
\usepackage{newtxtext,newtxmath}
\usepackage[T1]{fontenc}

\usepackage{graphicx,graphics}	
\usepackage{color,epsfig}
\usepackage{amsmath}	
\usepackage{amssymb}	
\usepackage{psfrag}
\usepackage[T1]{fontenc}
\usepackage{ebgaramond}
\usepackage{newtxmath}

\def\n{\hat{\bm{n}}}

\newcommand{\bfth}{\boldsymbol{\bf{\theta}}}

\newcommand{\bfl}{\boldsymbol{\ell}}

\newcommand{\Hi}{H\,\textsc{i}}
\newcommand{\HI}{H\,\textsc{i}~}

\def\k{{\bf k}}

\def\a{{\bm{a}}}
\def\x{{\bm{x}}}
\def\r{{\bm{r}}}
\def\U{{\bm{U}}}

\def\v{\bm{v}}

\def\V{\mathcal{V}}
\def\N{\mathcal{N}}
\def\S{{\mathcal S}}
\def\u{{\bf U}}

\begin{document}

\title{A visibility-based angular bispectrum estimator for radio-interferometric data}

\correspondingauthor{Sukhdeep Singh Gill}
\email{sukhdeepsingh5ab@gmail.com}

\author[0000-0003-1629-3357]{Sukhdeep Singh Gill}
\affiliation{Department of Physics, Indian Institute of Technology Kharagpur, Kharagpur 721 302, India}

\author[0000-0002-2350-3669]{Somnath Bharadwaj}
\affiliation{Department of Physics, Indian Institute of Technology Kharagpur, Kharagpur 721 302, India}

\author[0009-0006-4121-4955]{Sk. Saiyad Ali}
\affiliation{Department of Physics, Jadavpur University, Kolkata 700032, India}

\author[0000-0003-1206-8689]{Khandakar Md Asif Elahi}
\affiliation{Centre for Strings, Gravitation and Cosmology, Department of Physics, Indian Institute of Technology Madras, Chennai 600036, India}

\begin{abstract}
Considering radio-interferometric observations, we present a fast and efficient estimator to compute the binned angular bispectrum (ABS) from gridded visibility data. The estimator makes use of Fast Fourier Transform (FFT) techniques to compute the bispectrum covering all possible triangle shapes and sizes.
Here, we present the formalism of the estimator and validate it using simulated visibility data for the Murchison Widefield Array (MWA) observations at $\nu=154.25$ MHz. We find that our estimator is able to faithfully recover the ABS  of the simulated sky signal with $\approx 10-15 \%$  accuracy for a wide variety of triangle shapes and sizes across the range of angular multipoles $46 \le \ell \le 1320$. In future work, we plan to apply this to actual data and also generalize it to estimate the three-dimensional redshifted 21-cm bispectrum. 
\end{abstract}





\section{Introduction}
There is considerable motivation to quantify the statistics of the sky signal at radio-wavelengths.
Radio interferometric observations of the redshifted 21-cm signal from neutral hydrogen (\Hi\relax) hold the potential to probe a wide range of cosmological and astrophysical phenomena in a large redshift range \citep{BA5}. In particular, 
several radio-interferometers, such as 
MWA\footnote{\url{https://www.mwatelescope.org/}} \citep{tingay13}, LOFAR\footnote{\url{https://www.astron.nl/telescopes/lofar/}} \citep{haarlem}, 
and HERA\footnote{\url{https://reionization.org/}} \citep{DeBoer_2017} are currently involved in efforts to detect the Epoch of Reionization (EoR) redshifted 21-cm signal in the frequency range $100-200 \, {\rm MHz}$. 
Much of the observational effort so far has focused on the power spectrum (PS).    Despite continued efforts, the EoR 21-cm  PS remains to be detected, and the lowest upper limit at present is   $\Delta^2(k) < (30.76)^{2} \, {\rm mK}^2$  at $k = 0.192\, h\, {\rm Mpc}^{-1}$  for  $z = 7.9$ from the HERA \citep{Abdurashidova2022}.

The PS, which quantifies the square of the amplitude of different Fourier modes, is adequate if the signal is a Gaussian random field that is completely quantified by its second-order statistics. However, the EoR 21-cm signal is predicted to be highly non-Gaussian \citep{Bharadwaj2005}. The PS completely misses the correlations between the phases of different modes. It is also oblivious to the geometry and topology (e.g., \citealt{Bag_2018,Bag_2019}) of structures in the non-Gaussian field. 
The bispectrum (BS) is the lowest-order statistic that is sensitive to the non-Gaussianity. A measurement of the EoR 21-cm BS has the potential to capture considerable information that is missed by the PS \citep{Bharadwaj2005}. 
Considerable effort has been made to predict the BS using simulations \citep{suamnm2018,watk2019,sumanm2020,kamran2021,gill_eormulti}. Several studies 
show that the BS undergoes a sign change as the topology of the \HI field evolves.
 The first sign flip occurs at the early stage of the EoR, which serves as a valuable indicator of the emergence of distinct ionized bubbles in the neutral background \citep{suamnm2018}. The second sign change, which occurs at the end stage of the EoR, indicates a further topological shift, where isolated \HI islands emerge within a largely ionized background \citep{raste_2023, gill_eormulti}. Furthermore, the quadrupole moment of the BS is sensitive to the model of the EoR, and it holds the potential to distinguish between inside-out and outside-in scenarios of the EoR \citep{gill_eormulti}. The inclusion of the 21-cm BS along with the PS also improves constraints on the EoR model parameters \citep{shiam2017,Watkinson_2022}. 
There is also considerable motivation to measure the 21-cm BS from other cosmological epochs like the Dark Ages \citep{pillep2007,cooray2008} and the post-reionization era \citep{ali2006,dsarkar2019}.

 \citet{trott2019} have estimated the 21-cm bispectrum from data taken as part of the EoR project of the MWA. In their study, they have considered equilateral and isosceles triangles, and found that the thermal noise level is achieved in $10$ h of observations for the case of large-scale isosceles triangles. They propose that the BS may be detectable with lesser observational time than the PS for radio-interferometers with dense $uv$ coverage. To the best of our knowledge, this is the only observational attempt to measure the 21-cm BS. 

In this paper we present the first step towards systematically developing a fast BS estimator considering triangles of all possible shapes and sizes. The analysis here is restricted to a single frequency, and we entirely focus on estimating the two-dimensional (2D) angular bispectrum (ABS). We plan to consider multi-frequency observations and the three-dimensional (3D)  BS in future work. The idea is to proceed in two steps, similar to the approach adopted earlier for the PS  where  \citet{samir14} considered the angular power spectrum (APS)  at a single frequency,  and \citet{Bh18} extended this to the 3D PS considering multi-frequency observations. 

It is worth noting that the ABS is of considerable interest in its own right. 
For example, there have been several efforts to probe the APS of the 
$\sim 150 \, {\rm MHz}$ sky signal using radio-interferometers, both to characterize the foregrounds for the 21-cm PS measurements and to study the diffuse Galactic synchrotron emission e.g., \citep{Ali2008, Bernardi2009,Ghosh2012}.   There is also considerable interest in measuring the APS of 21-cm emission from \HI in the interstellar medium of galaxies \citep{begum2006,dutta2009}, and also the continuum emission from ionized gas in supernova remnants \citep{Roy_2008,Saha19,saha2021}. In all of these contexts, it would be very interesting to enhance our knowledge by including studies of the ABS. 

In this work, we present a visibility-based estimator for the ABS
 ${B}(\ell_1, \ell_2, \ell_3)$. The estimator uses gridded visibilities to compute the binned ABS using a fast FFT-based technique \citep{Sefusatti_thesis,Jeong_thesis,sco2015} considering triangles of 
  all possible shapes and sizes. In this work, we present the formalism of the estimator and validate it using simulated MWA observations at $\nu=154.25$ MHz.

The paper is organized as follows. In Section \ref{sec:meth}, we present the mathematical formalism of the estimator, and in Section \ref{validation} we discuss our method for validating the estimator. We present the results in Section \ref{sec:results},  whereas 
we present a summary and conclusions in Section \ref{sec:sum}. 

\section{Bispectrum Estimator}
\label{sec:meth}
We consider the brightness temperature fluctuations $\delta T_b(\bfth)$ from a region of the sky that subtends a solid angle $\Omega \ll 1$ that is sufficiently small so that it may be approximated as a two-dimensional (2D) plane. We also express this in terms of  Fourier components $\Delta \Tilde{T}_b(\bfl)$,
\begin{equation}
\delta T_b(\bfth) = \Omega^{-1}\sum\limits_{\bfl} \exp[ -i \bfl \cdot \, \bfth]\,\Delta \Tilde{T}_b(\bfl)~ \, , 
\label{eq:TbFT}
\end{equation}
where $\bfl$, which is the Fourier conjugate of $\bfth$, may also be interpreted in terms of  the angular multipole $\ell=\mid \bfl \mid$. 
Note that the entire analysis here is restricted to a single frequency $\nu$ (and wavelength $\lambda$) which we do not show explicitly as an argument. 
Considering $\delta T_b(\bfth)$ to be a statistically homogeneous and isotropic random field, we have the angular power spectrum (APS) 
\begin{equation}
C_{{\ell}}=\Omega^{-1} \langle~ \Delta \Tilde{T}_b({\bfl}) \, \Delta  \Tilde{T}^*_b({\bfl})~\rangle 
\label{eq:maps1}
\end{equation}
and the angular bispectrum (ABS) 
\begin{equation}
B (\ell_1, \ell_2, \ell_3)=\Omega^{-1}\langle~ \Delta \Tilde{T}_b(\bfl_1) \Delta\Tilde{T}_b(\bfl_2) \Delta \Tilde{T}_b(\bfl_3)~\rangle  ~\,,
 \label{eq:mabs1}
\end{equation}
where   $\bfl_1+ \bfl_2+ \bfl_3=0$ {\it i.e.} they form
a closed triangle, and the angular brackets $\langle... \rangle$ denote an ensemble average over independent realizations of the random field. Note that $C_{{\ell}}$ only depends on magnitude  $\ell= \mid\bfl\mid$, and $B (\ell_1, \ell_2, \ell_3)$ only depends on the shape and size of the triangle which is entirely specified by $(\ell_1, \ell_2, \ell_3)$ the 
lengths of the three sides respectively. Considering $\ell_1\geq \ell_2\geq \ell_3$, here we use 
$\ell_1$ to quantify the size, and the dimensionless parameters $ \mu =- {(\bfl_1 \cdot \bfl_2)}/{(\ell_1 \ell_2)}$ and $t={\ell_2}/{\ell_1}$ to quantify the shape of the triangle. 
The allowed parameter values are constrained to the range $0.5 \le \mu,t \le 1$  with $2 \mu t \ge 1$, and the reader is referred to \citet{bharad2020} for a detailed discussion of this parametrization of the bispectrum $B(\ell_1,\mu,t)$.

 The $\V(\u)$ visibilities measured in radio interferometric observations are a sum of the sky signal and the system noise. 
 In the present work, we have focused on the sky signal and ignored the system noise. As mentioned earlier, we have adopted the flat-sky approximation which treats the region of the sky under observation as a flat 2D plane.
 We further assume a coplanar radio interferometric array pointing vertically upwards. We then have
\begin{equation}
\V(\u)= Q\, \int \, d^2 \theta\, {A}(\bfth) ~\delta T(\bfth) ~\exp[{i2\pi\u\cdot\bfth}] 
\label{eq:vis}
\end{equation}
 where the 2D vector  $\u$, with components $(u,v)$, is a baseline, $Q$ is the Rayleigh-Jeans factor $Q=2 k_B/\lambda^2$ conversion factor from brightness temperature to specific intensity, and  ${A}(\bfth)$ is the antenna primary beam pattern, which typically is not known a priori. Although this can be estimated from observations \citep{Line_2018,hera_beam_2020,lofar_beam_2022}, it is often useful to consider simple models for ${A}(\bfth)$ (e.g., \citealp{Bharadwaj_2001,Choudhuri2014}). 
 For the work presented here, we consider the  Murchison Widefield Array  (MWA, \citealp{Wayth_2018}) 
 at $\nu=154.25$ MHz.  It is possible to model each MWA tile as a square aperture of $d=4$m, whereby   \citep{Line_2018,chatterjee_2023}
\begin{equation}
    {A}(\bfth) = \text{sinc}^2\bigg(  \dfrac{\pi d\theta_x}{\lambda}\bigg)~ \text{sinc}^2\bigg(  \dfrac{\pi d\theta_y}{\lambda}\bigg) \,,
    \label{eq:beam}
\end{equation}
for which the  FWHM of the primary beam is $\theta_{F}=24.68\deg$.

Following \citet{Choudhuri2014},  it is useful to approximate  ${A}(\bfth)$ as Gaussian     to analytically compute the  relations between visibility correlations and the statistics of the sky signal, and we use 
\begin{equation}\label{eq:A_G}
    {A}_G(\bfth) = \exp \left [{-\theta^2/\theta_0^2}\right]
\end{equation}
where $\theta_0=0.6\theta_F$. This approximation holds well at small angles, within $\theta \le \theta_F$.
At large baselines $(U \gg 1/\theta_0)$, we then have (see appendix \ref{app:2vis_derivation} for a derivation)
\begin{equation}
\langle  \V(\u){\V}^*(\u+\Delta\u) \rangle \, =  \dfrac{\pi\theta_0^2 Q^2}{2} \exp\left [{-\pi^2\,\theta_0^2\, \Delta U^2/2}\right]\, C_{\ell}\,  
\label{eq:2vis}
\end{equation}
which relates the two visibility correlation to the  APS $C_{\ell}$   with $\ell=2 \pi U$. 
We do not expect Eq.~(\ref{eq:2vis}) to hold at small baselines $U \le 1/\theta_0$, where it is necessary to consider the convolution of $C_{\ell}$ with the Fourier transform of $ {A}(\bfth)$. 
The reader is referred to 
\citet{Choudhuri2014}  for a detailed discussion of the derivation and application of this relation. 

The three visibility correlation is similarly related to the ABS, and at large baselines we have (see appendix \ref{app:2vis_derivation} for a derivation)
\begin{equation}
\begin{aligned}
\langle \V(\u_1) & \V(\u_2) \V(\u_3+\Delta \u)\rangle \\&=  \dfrac{\pi\theta_0^2 Q^3}{3} \exp\left [{-\pi^2\theta_0^2\Delta U^2/3} \right ] B({\ell_1}, {\ell_2}, {\ell_3}) \,.
\label{eq:S3}
\end{aligned}
\end{equation}
We do not expect this equation to hold at small baselines ($U \le 1/\theta_0$), where it is necessary to consider the convolution of $B({\ell_1}, {\ell_2}, {\ell_3})$  with the Fourier transform of $ {A}(\bfth)$.   Here, $\u_1 + \u_2 + \u_3=0$ forms a closed triangle, and $\Delta \u$ is the deviation from a closed triangle configuration. We see that the correlation is strongest when $\Delta U=0$, and it falls off rapidly as $\Delta U$ increases.
There is negligible correlation for $\Delta U \ge (\pi \theta_0)^{-1}$. 
Here, we have used Eq.~(\ref{eq:S3}) to define a visibility-based binned angular bispectrum estimator.

To proceed further, we introduce a square grid of spacing $\Delta U_g$ in the $(u,v)$ plane, and assign each visibility $\V(\u_i)$   to the grid point $\u_g$ nearest to $\u_i$ using 
\begin{equation}
\V_{g} = \sum^{N_g}_{i}\tilde{w}(\u_g-\u_i) \, \V(\u_i) \, ,
\label{eq:vg}
\end{equation}
and use $N_g$ to denote the number of visibilities contributing at any grid point $g$. 

\begin{figure}
\centering
\includegraphics[width=.49\textwidth]{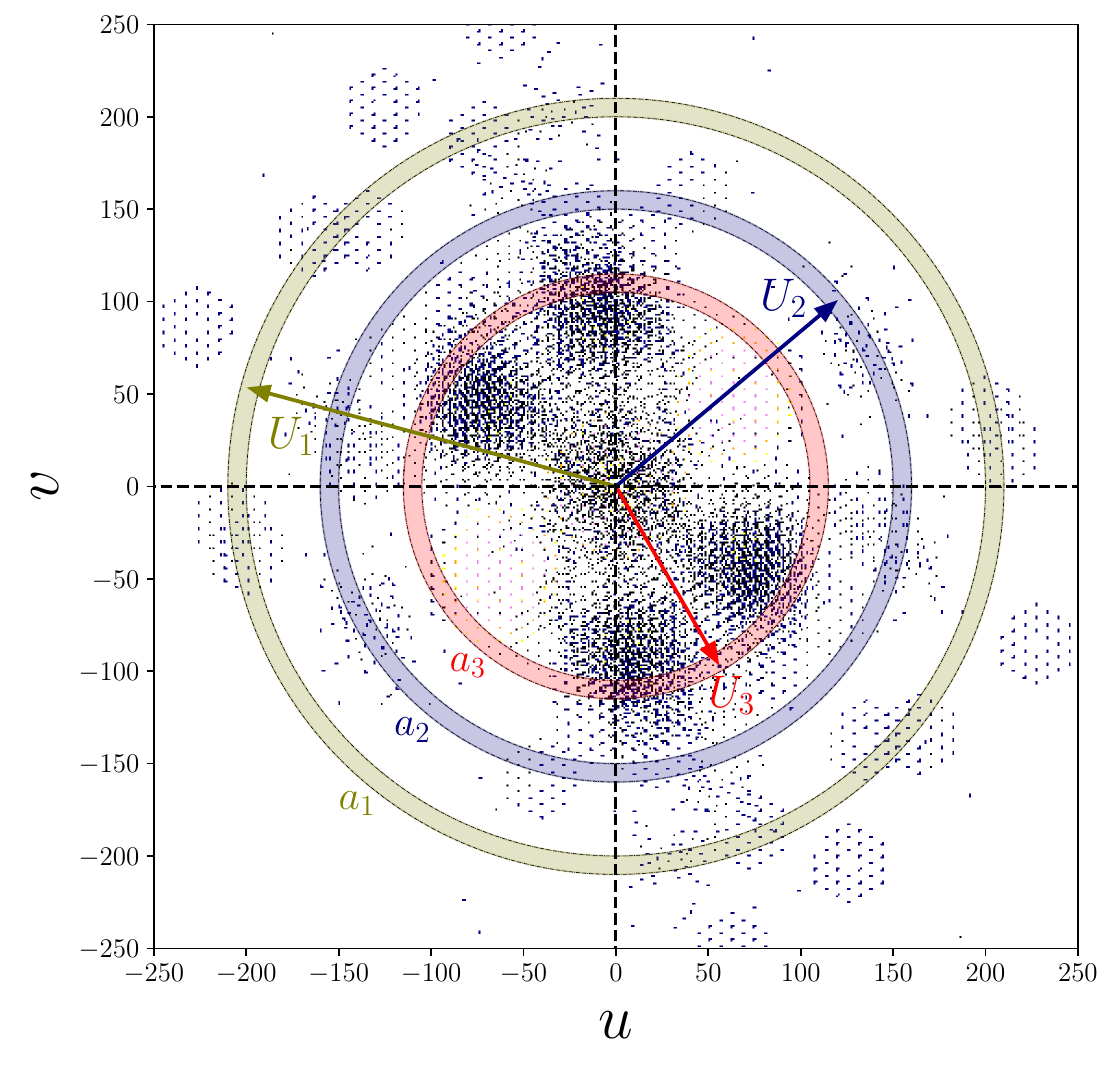}
\caption{The binning scheme of the estimator. The scattered dots show the discrete sampling of ($u,v$) space (gridded baseline distribution $\u_g$) corresponding to a particular pointing of the drift scan observation of the MWA telescope at $\nu=154.25$ MHz. The $\u$ space is divided into several annular rings. Three such rings (labeled as $a_1,a_2,a_3$) with average radius ($U_1,U_2,U_3$) are shown here schematically. The combination of these three rings corresponds to a set of triangles having unique shapes and sizes, and defines a single bin of triangles. One such triangle formed by three discrete modes $\u_{g1}+\u_{g2}+\u_{g3}=0$ is shown.}
\label{fig:binV}
\end{figure}

For the present work, we have considered the baseline distribution corresponding to a particular pointing of the drift scan observations presented in \citet{Patwa_2021} and also analyzed in \citet{chatterjee_2023} and \citet{chatterjee_2024}.  Fig.\ref{fig:binV} shows $N_g$ corresponding to the gridded visibilities for this data. 
Here, we use the gridded visibilities $\V_{g}$ to estimate the bispectrum. The computation scales as $\sim N_t^4$, where $N_t$ is the total number of grids, if we evaluate this directly by correlating all possible triplets of grid points that form a closed triangle.  
 The computation can be reduced to $\sim N_t^2\log{N_t^2}$ by utilizing FFT techniques introduced by \citep{sefu2006, sco2015}. Here we follow  \cite{shaw21} to present an  FFT  based fast estimator for the binned angular bispectrum $B(\ell_1,\mu,t)$.

We divide the $\u$ plane into annular rings. Three such rings,  labeled  $(a_1,a_2,a_3) $ 
with  mean radii $(U_1,U_2,U_3)$  respectively,  are illustrated in  Fig. \ref{fig:binV}.
Considering any ring $a_m$, we define 
\begin{equation}
    D(\ell_m,\bfth)=\sum_{g\in a_m} W_{g} \V_g \exp(-i  \bfl_{g} \cdot\bfth)~, 
    \label{eq:iFT}
\end{equation} 
which is the inverse Fourier Transform of $\V_g$ restricted to the annular ring $a_m$, with  
$\bfl_{g}=2 \pi \u_g$ and $\ell_m=2 \pi U_m$. Note that we have introduced weights $W_g$ for the 
gridded visibilities $\V_g$, these can be adjusted to optimize the signal to noise ratio of the 
estimated bispectrum.  Here, we have not included the system noise contribution, and we use $W_g=N_g^{-1}$  and $0$ for the filled and empty grids, respectively. 
We similarly define 
\begin{equation}
    I(\ell_m,\bfth)=\sum_{g\in a_m}  W_g \exp(-i\bfl_{g} \cdot\bfth)~. 
    \label{eq:iFT_I}
\end{equation} 
which is the inverse Fourier Transform of $W_g$ restricted to the annular ring $a_m$.  
Following   \cite{shaw21}, we use a combination of three rings $(a_1,a_2,a_3)$ with $\ell_1 \ge \ell_2 \ge \ell_3$, to define the binned angular bispectrum estimator 
\begin{equation}
\Hat{B}(\ell_1,\ell_2,\ell_3)
= \dfrac{1}{A}  \dfrac{\sum\limits_{\bfth} D(\ell_1,\bfth) D(\ell_2,\bfth)  D(\ell_3,\bfth)}{\sum\limits_{\bfth} I(\ell_1,\bfth) I(\ell_2,\bfth)  I(\ell_3,\bfth)} ~.
\label{eq:estimator_t}
\end{equation}
where $A=\pi\theta_0^2/3$ is a normalization constant (Eq.~\ref{eq:S3}). 

The estimator considers all closed triangles $\u_{g1}+\u_{g2}+\u_{g3}=0$ such that 
$(\u_{g1},\u_{g2},\u_{g3})$ are within the annular rings $(a_1,a_2,a_3)$ respectively, and it provides the weighted average over all such triangles. One such triangle is illustrated in Fig. \ref{fig:binV}. We refer to this weighted average (Eq.~\ref{eq:estimator_t}) as the binned angular bispectrum estimator $\hat{B}(\ell_1,\mu,t)$ where $\ell_1$ quantifies the average size of the triangles in the bin, and the parameters $\mu=(\ell_1^2+\ell_2^2-\ell_3^2)/(2 \ell_1 \ell_2)$ and $t=\ell_2/\ell_1$ together quantify the average shape of the triangles in the bin. 

We note that the system noise contribution to the different visibilities are uncorrelated, and we have  ${B}(\ell_1,\mu,t)= \langle \hat{B}(\ell_1,\mu,t) \rangle \equiv \langle \hat{B}(\ell_1,\ell_2,\ell_3) \rangle$ even if the system noise contribution is taken into account. 
However, the system noise will make an extra contribution to the statistical fluctuations in the estimated ABS.

\section{Validating the estimator}
\label{validation}
We validate the estimator by simulating a non-Gaussian sky signal $\delta T_{\rm b} (\bfth)$ for which the analytical expression for the ABS is known. We start from a Gaussian random field $\delta T_{\rm G} (\bfth)$ generated using an input model APS $C_{\ell}=(\ell/\ell_0)^{-1}\, \exp[-\ell^2/(\pi \ell_0)^2] \,  {\rm mK^2}$   with $\ell_0= 1000$.  The non-Gaussian random field $\delta T_{\rm b} (\bfth)$ is obtained using a local non-linear transformation,
\begin{equation}
\delta T_{\rm b} (\bfth)= \delta T_{\rm G} (\bfth) + \dfrac{f_{\rm NG}}{\sigma_T}\,(\delta T^2_{G} (\bfth)- \sigma_{ T}^2) \,, 
\label{eq:mb1}
\end{equation}
where the dimensionless parameter $f_{\rm NG}$ controls the level of non-Gaussianity,
 and $\sigma_{ T}$ is the standard deviation of $\delta T_{\rm G} (\bfth)$. The analytical expression for the ABS of $\delta T_{\rm b} (\bfth)$ calculated to the first order in $f_{\rm NG}$ is (see appendix \ref{app:bispectrum_derivation} for the details),
\begin{equation}
B_{\rm Ana}(\ell_1,\ell_2, \ell_3) = \dfrac{2\,f_{\rm NG}}{\sigma_T} \Big({C_{\ell_1}}\,{C_{\ell_2}}+{C_{\ell_2}}\,{C_{\ell_3}}+ {C_{\ell_3}}\,{C_{\ell_1}}\Big) \,,
\label{eq:bsana}
\end{equation}
which is valid for $f_{\rm NG}\ll1$. Here we have used $f_{\rm NG}=0.17$, for which the ABS estimated from the simulated sky signal was found to be consistent with the predictions of Eq.~(\ref{eq:bsana}). 

The MWA baselines, for the data considered here,  are mostly  $(\sim 99 \%)$   within $U =250$ \citep{chatterjee_2023} that corresponds to $\ell =1570$, which is an angular scale of $0.115^{\circ}$. Here, we have simulated the sky signal on a flat  2D grid of spacing  $0.029^{\circ}$, which spans  $117.35^{\circ}$ that is $\sim 4.5$ times larger than $\theta_{F}$. We have chosen this large range of angles to avoid abruptly cutting off the simulated signal at either the small angular scales or the large angular scales.  The Gaussian factor in $C_{\ell}$ smoothly cuts off the signal at the smallest angular scales on the grid. We have multiplied the simulated  $\delta T_{\rm b} (\bfth)$  with $ {A}(\bfth)$ (Eq.~\ref{eq:beam}), and used a DFT (Eq.~\ref{eq:vis}) to calculate the  simulated  $\V(\u)$.  The primary beam pattern $ {A}(\bfth)$  smoothly cuts off the sky signal from the largest angular scales $(\theta > \theta_F)$ on the grid. In principle, we should use a spherical sky to perform these simulations (e.g., \citealp{chatterjee_2023}), however, the simulations are significantly faster if we use the flat sky approximation adopted here. Further, the subsequent analysis is restricted to the range $20 \le \ell \le 1570$ where we expect the flat sky approximation to hold \citep{Datta2007}.

We have gridded the visibilities (Eq.~\ref{eq:vg}) using a grid spacing of $\Delta U_g= \sqrt{\ln2}/{\pi\theta_0}\approx 1$. Here, instead of correlating the visibilities $\V(\u)$ at three baselines that form a closed triangle (Eq.~\ref{eq:S3}), we estimate the ABS by correlating  $\V_g$ at three grid points that form a closed triangle. The different baselines that contribute to the three grid points generally do not form closed triangles. 
The relatively small grid spacing used here ensures that the factor $e^{-(\pi^2\theta_0^2\Delta U^2/3)}$ that arises in Eq.~(\ref{eq:S3}) 
due to this does not fall much below $1$. We find that this factor has a value of $0.89$ for a typical value of $(\Delta U)^2=(\Delta U_g)^2/2$.

We have divided the $(u,v)$ plane (Fig.~\ref{fig:binV}) 
 into $22$ concentric annular rings of varying width, with a single ring of width 4  spanning radius 1 to 5 (in grid units), 
  nine equally spaced rings between radii $5$ and $50$, five between 50 and 100, five between 100 and 200, and two between 200 and 250. We have used Eq.~(\ref{eq:estimator_t}) to estimate  $B(\ell_1,\mu,t)$ for every possible combination of three annular rings.  
  Each estimated $\hat{B}(\ell_1,\mu,t)$ corresponds to the average ABS for all possible closed triangles that have one $\u$ respectively in each of the three annular rings, as illustrated in Fig.~\ref{fig:binV}.  The bin widths have been progressively increased with $U$ to account for the fact that the expected signal $B(\ell_1,\mu,t)$ and the baseline number density both go down with increasing $U$.   Choosing a larger number of finer rings would provide estimates of the ABS at small intervals of $(\ell_1,\mu,t)$, at the cost of increasing the computation time, whereas choosing a smaller number of coarser rings would have the opposite effect. 
  For the choice of rings adopted here, the different estimates of $\hat{B}(\ell_1,\mu,t)$ do not occur at equal intervals in the $(\ell_1,\mu,t)$ parameter space. We have divided the $\ell_1$ range into bins of equal logarithmic spacing, and the $\mu$ and $t$ ranges into bins of equal linear spacing, and averaged  (weighted by the number of triangles) the $\hat{B}(\ell_1,\mu,t)$ values that occur in each bin. 
  
  We have used $500$ statistically independent realizations of the random field to obtain reliable estimates of the ensemble average  ${B}(\ell_1,\mu,t) =\langle \hat{B}(\ell_1,\mu,t) \rangle$ and the variance $\sigma^2=  \langle [\hat{B}(\ell_1,\mu,t)]^2 \rangle - [{B}(\ell_1,\mu,t)]^2$, for which the results are presented in the next section.

\section{Results}
\label{sec:results}

 \begin{figure}
\centering
\includegraphics[width=.48\textwidth]{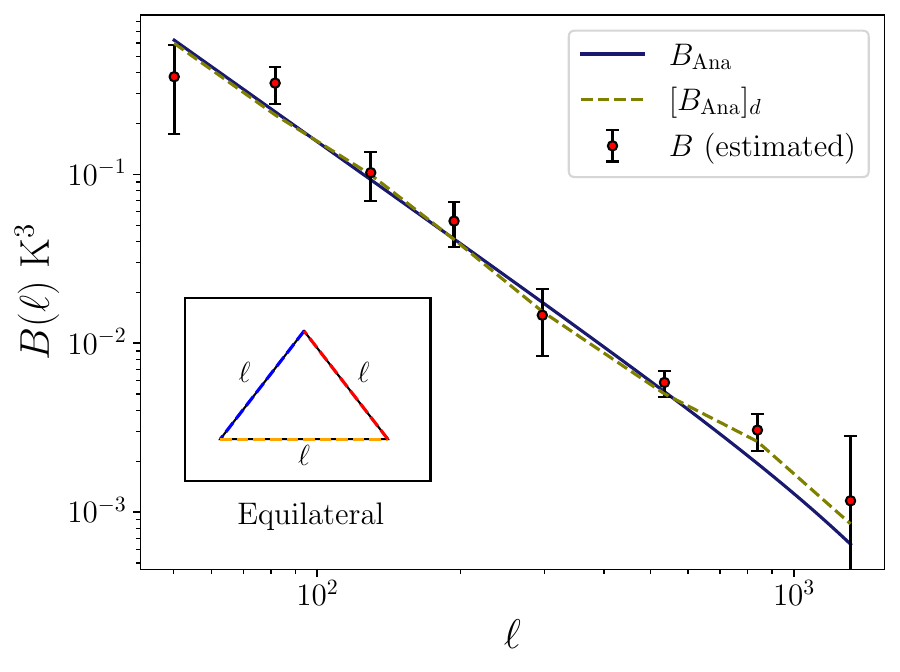}
\caption{Validation of the estimator. Results are shown for the equilateral triangle at various angular multipoles  $\ell$. The red circles show the estimated bispectrum from simulated MWA visibility data, and the error bars show r.m.s. statistical fluctuations of the estimates computed using 500 independent realizations. The blue solid line shows the analytical predictions (Eq.\ref{eq:bsana}). The green dashed line shows the analytical prediction computed by incorporating the discrete sampling of the $\ell$ modes available in the data for each bin.}
\label{fig:equi}
\end{figure}

Fig. \ref{fig:equi} shows the results considering  equilateral triangles ($\mu\approx0.5,t\approx1$), for which the ABS is predicted to be  $B_{\rm Ana}(\ell) = 6\,(f_{\rm NG}/{\sigma_T}) \,{C_{\ell}}^2\,$ (Eq.~\ref{eq:bsana}) which is represented by the blue solid line. The red circles show the binned ABS ${B}(\ell)$ estimated from the simulated visibilities, whereas the error bars show the $1\sigma$ r.m.s. statistical fluctuations.
Each $B(\ell)$ shown here corresponds to the average of values estimated at some discrete $\ell$, the exact set of $\ell$ values available within each bin depends on the choice of annular rings (Fig.~\ref{fig:binV}).  To account for this, we also show the analytic prediction $[B_{\rm Ana}]_d$ (green dashed line) which incorporates the same discrete sampling as the actual data.  We see that  $B_{\rm Ana}(\ell)$ and $[B_{\rm Ana}]_d$ are nearly indistinguishable for $\ell<600 \, (U < 100)$, however, we have a noticeable (but relatively small) difference at larger $\ell$ where we have used wider annular rings.   We find that the estimated  ${B}(\ell)$  are  are consistent with $[B_{\rm Ana}]_d$, and the deviations are within $\pm 1 \sigma$ for  for    $\ell \ge 80$. The deviations are within $ \pm 2\sigma$ for $\ell < 80$, which may also be interpreted as arisen due to statistical fluctuations.    The estimated  ${B}(\ell)$ are all  consistent with $B_{\rm Ana}$ also within $2\sigma$.
There is one more estimate at $\ell\approx 20$, which deviates significantly due to the convolution with the primary beam, and we have not shown this here.

 \begin{figure*}
    \centering 
\includegraphics[width=.99\textwidth]{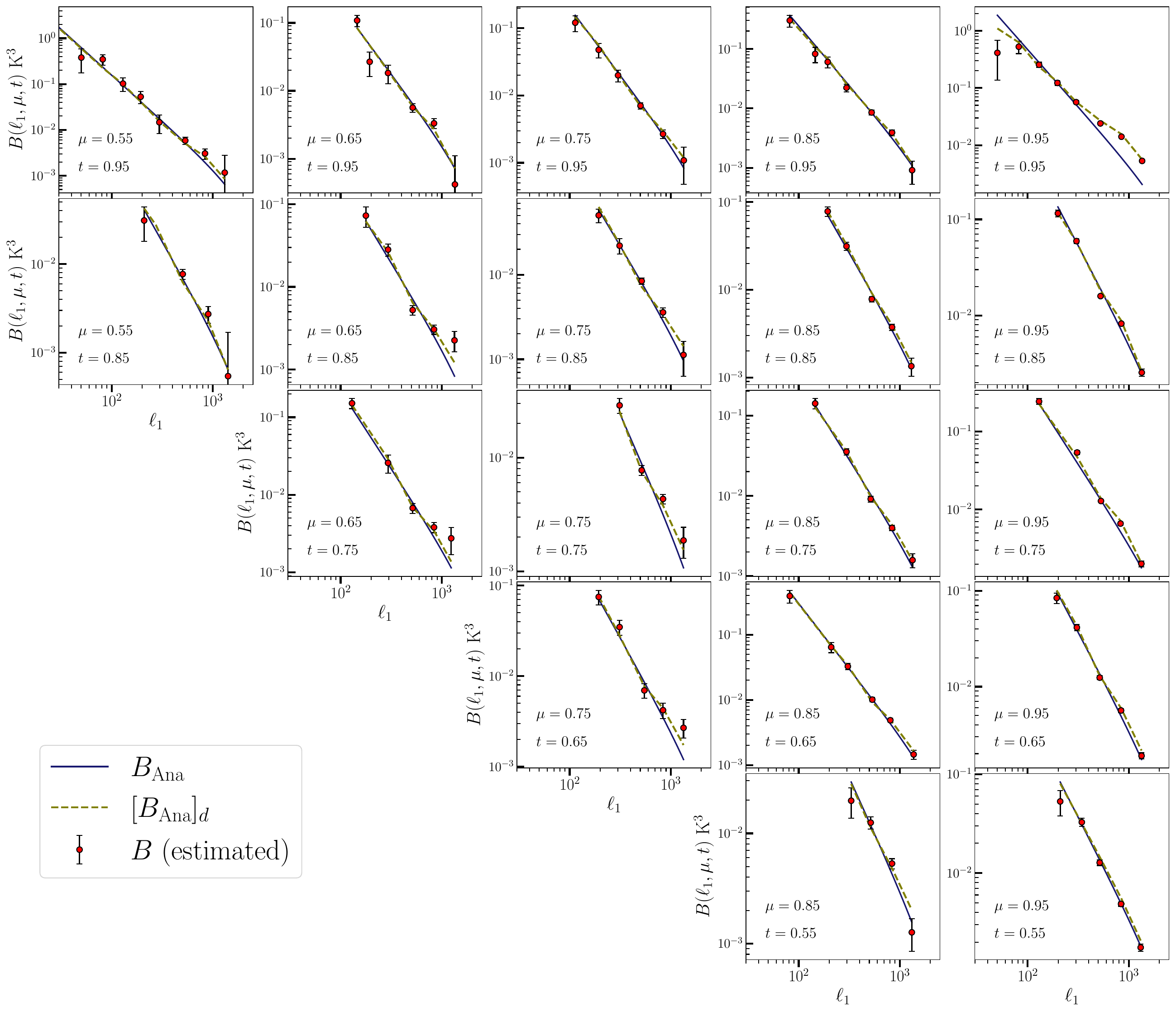} 
    \caption{The upper left panel, which shows $B(\ell_1,\mu,t)$ as a function of $\ell_1$ for $(\mu,t)=(0.55,0.95)$ fixed, is exactly the same as Fig. \ref{fig:equi} that considers equilateral triangles. The other panels are similar, but each corresponds to different values of $(\mu,t)$, which corresponds to a different triangle shape. The panels span the entire allowed range of $(\mu,t)$ values, covering triangles of all possible shapes (refer to Fig. 2 of \citealp{bharad2020}). }
    \label{fig:all_tri}
\end{figure*}

Fig.~\ref{fig:all_tri} provides a comprehensive validation of the estimator, considering triangles of all possible shapes $(\mu,t)$ and sizes $\ell_1$. Each panel of Fig.~\ref{fig:all_tri} is equivalent to Fig.~\ref{fig:equi}, but for a different triangle shape. Each panel  shows $B(\ell_1,\mu,t)$  as a function of $\ell_1$ for a fixed set of $(\mu,t)$. We have divided the allowed range  $0.5 \le \mu,t \le 1$ into $5\times5$ linear bins, and show results only for the bins that fall within the allowed region $ 2 \mu t  \ge 1 $.
 The reader is referred to Fig.~2 of \cite{bharad2020} for a detailed discussion of 
 the location of various triangle shapes in the $\mu-t$ plane. The upper left corner corresponds to equilateral triangles, for which the results have already been presented in Fig.~\ref{fig:equi}.  The upper and lower boundaries, respectively, correspond to isosceles triangles of two different types, whereas the right boundary corresponds to linear triangles, where the three sides are aligned in nearly the same direction. The upper and lower right corners respectively correspond to squeezed $(\ell_1 \approx \ell_2, \ell_3 \rightarrow 0)$ and stretched  $(\ell_1/2 \approx  \ell_2 \approx \ell_3 )$ triangles. 

We see that the results shown in the different panels of Fig.~\ref{fig:all_tri} are broadly very similar to those shown in Fig.~\ref{fig:equi}, which has already been discussed in some details. However, there are also some differences that we highlight below. First, note that the available $\ell_1$ range depends on the shape of the triangle, and we have the maximum $\ell_1$ range for equilateral and squeezed triangles.  The exact $\ell_1$ range available for any shape is decided by the sizes of the annular rings that we have used. The relative sizes of the $1 \sigma$ error bars also show considerable variation. These are typically large for low $\ell_1$, and decrease with larger $\ell_1$.  The error bars also vary with the triangle shape, and these are smaller near the squeezed triangles.  Both of these features are related to the number of available triangles.  In general, we see that most values of $B(\ell_1,\mu,t)$ agree well with
$[B_{\rm Ana}]_d$, broadly validating our estimator.  The subsequent results provides a more quantitative comparison between $B(\ell_1,\mu,t)$ and  $[B_{\rm Ana}]_d(\ell_1,\mu,t)$.

  \begin{figure*}
    \centering 
    \includegraphics[width=.96\textwidth]{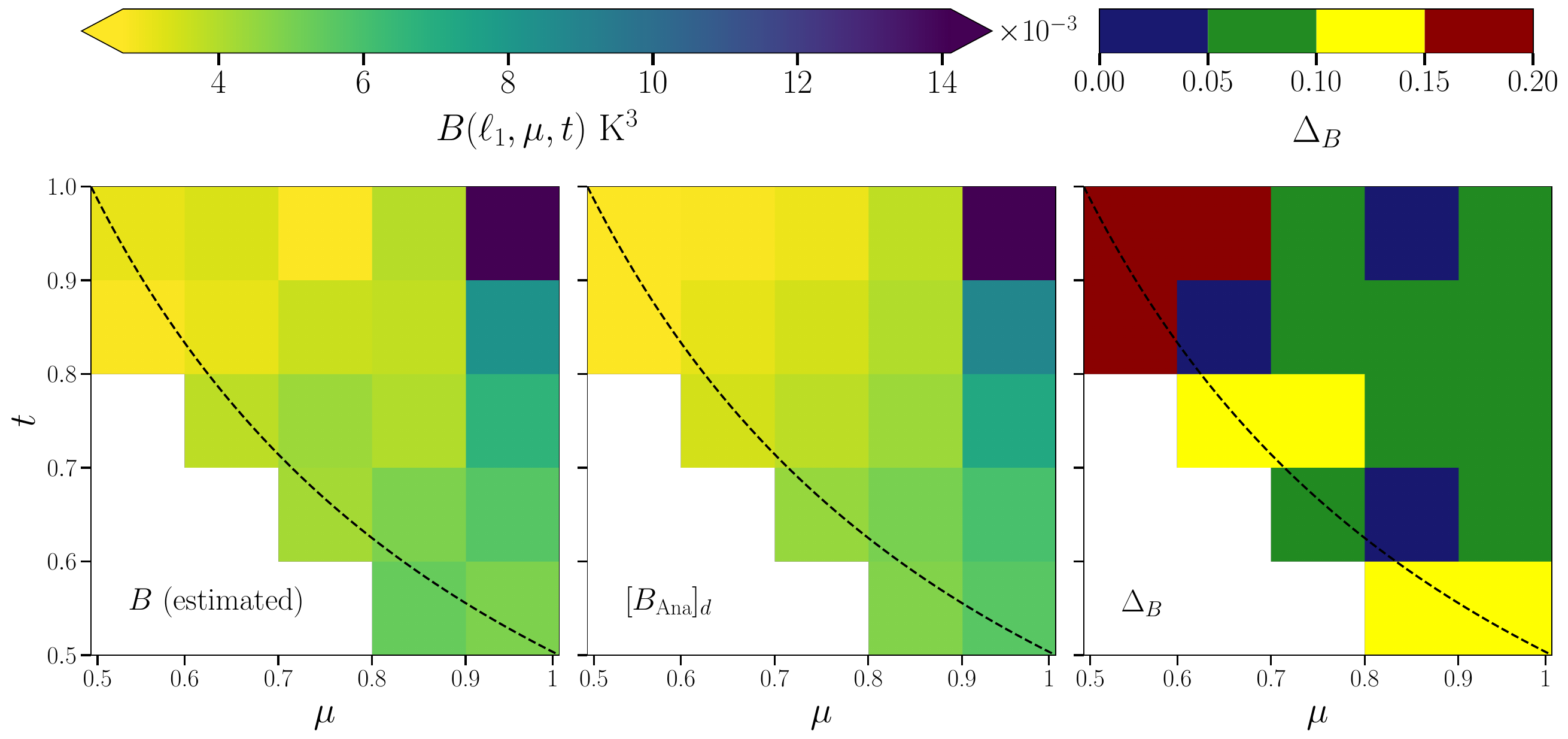} 
    \caption{ The left panel shows the estimated bispectrum $B(\ell_1,\mu,t)$
    as a function of $(\mu,t)$ at $\ell_1=819$ fixed. Here  $\mu$ and $t$ parameterize the shape of a triangle.  The allowed parameter range that is bounded within $2\,\mu\, t = 1$ (dashed line) and $0.5 \le \mu,t \le 1$ uniquely covers triangles of all possible shapes. The  
     middle panel shows  the analytic prediction  $[B_{\rm Ana}]_d$, and 
 the right panel shows $\Delta_B=|B-[B_{\rm Ana}]_d|/B$ the fractional deviations of the estimates from the predicted values. }
    \label{fig:l1_fix}
    \end{figure*}

The left panel of Fig.~\ref{fig:l1_fix} shows the $(\mu,t)$ dependence of $B(\ell_1,\mu,t)$, considering the fixed value $\ell_1=819$ for which all shapes are well sampled (Fig.~\ref{fig:all_tri}).  We see that the value of $B(\ell_1,\mu,t)$ is the maximum for squeezed triangles, which occurs in the upper right corner.  The value of $B(\ell_1,\mu,t)$ decreases relatively faster along $\mu$ as compared to $t$, and $B(\ell_1,\mu,t)$ is minimum for equilateral triangles. We have $B(\ell_1,\mu,t) \propto \ell_1^{-2}$, and we may expect a similar $(\mu,t)$ dependence for other $\ell_1$ also, barring effects due to the discrete sampling. The  middle panel shows $[B_{\rm Ana}]_d(\ell_1,\mu,t)$ for the same $\ell_1$ value. We see that  $B(\ell_1,\mu,t)$ and $[B_{\rm Ana}]_d(\ell_1,\mu,t)$ both show very similar behavior. The right panel shows the fractional deviation $\Delta_B=|B-[B_{\rm Ana}]_d|/[B_{\rm Ana}]_d$. We see that most of the values of $\Delta_B$ are well within $15\%$, except for three bins near the equilateral triangle where the fractional deviation has values $\approx 15-20\%$.

  \begin{figure*}
    \centering 
    \includegraphics[width=.98\textwidth]{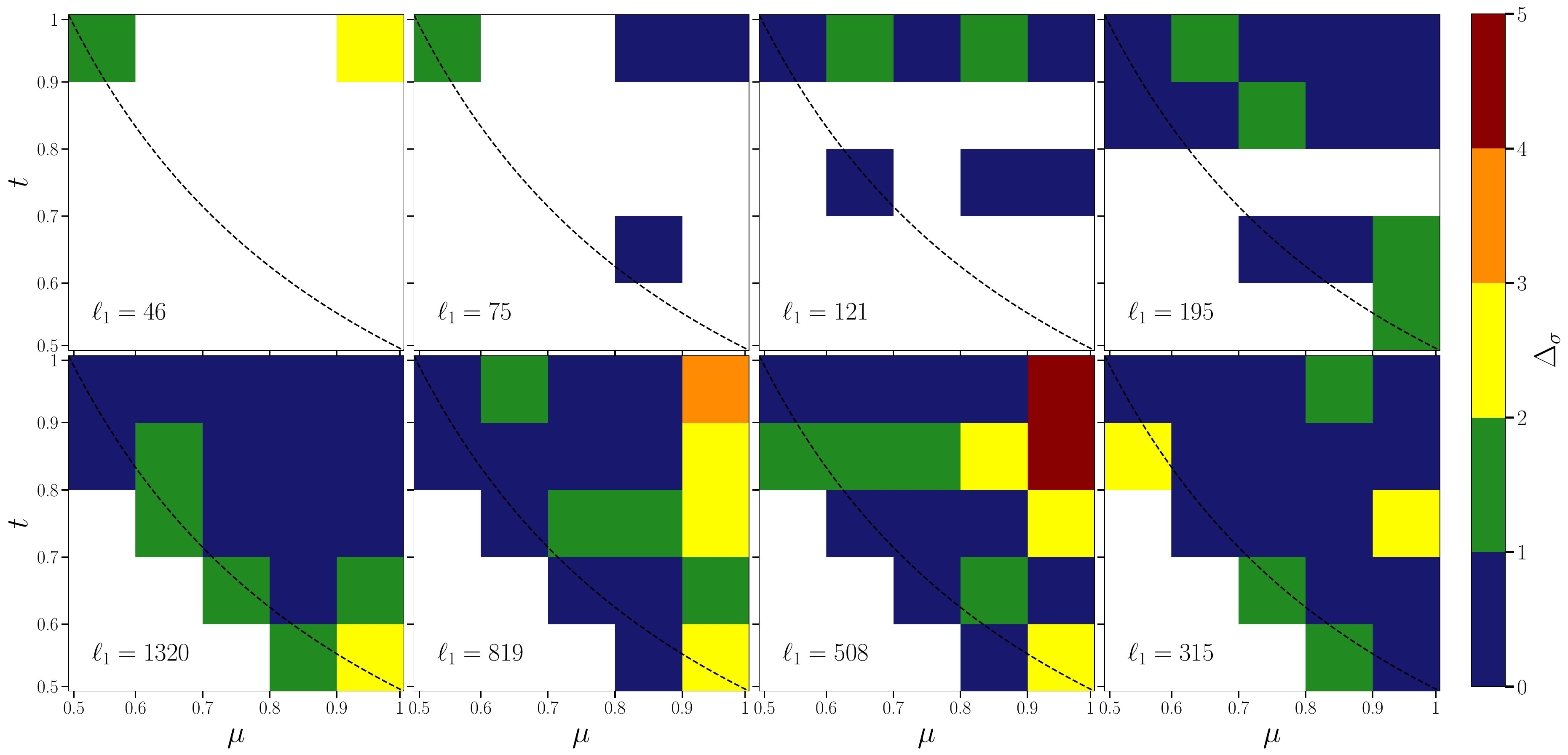} 
    \caption{ Each panel considers a different $\ell_1$,  for which it shows $\Delta_\sigma=|B-[B_{\rm Ana}]_d|/\sigma$ as a function of $(\mu,t)$. Here $\Delta_\sigma$ is the difference between the estimated bispectrum $(B)$ and the analytical prediction $([B_{\rm Ana}]_d)$, expressed in units of the expected statistical fluctuations $\sigma$.  The value of $\ell_1$ increases clockwise
     starting from the upper left panel. Here  
    $\mu$ and $t$  parameterize the shape of a triangle.  The allowed parameter range that is bounded within $2\,\mu\, t = 1$ (dashed line) and $0.5 \le \mu,t \le 1$ uniquely covers triangles of all possible shapes.  
  }
    \label{fig:sigma}
\end{figure*}

We have considered the ratio $\Delta_\sigma=|B-[B_{\rm Ana}]_d|/\sigma$
to analyze whether the deviations between the estimated values $B(\ell_1,\mu,t)$ and the predicted values $[B_{\rm Ana}]_d(\ell_1,\mu,t)$ are consistent with those expected from statistical fluctuations or not. 
 The different panels of  Fig~\ref{fig:sigma} correspond to different values of $\ell_1$, starting from $\ell_1=46$ in the upper left corner and increasing clockwise to $\ell_1=1320$ in the bottom left corner. Each panel shows $\Delta_\sigma$ as a function of $\mu$ and $t$. We do not show results for the smallest bin $\ell_1=22$ where we have only a single estimate at the equilateral triangle, for which $\Delta_\sigma \le 2$ is consistent with statistical fluctuations. Considering the smallest $\ell_1$, we have estimates for only two $(\mu,t)$ bins, namely the equilateral and squeezed triangles, for which we have $\Delta_\sigma \le 2 $ and $\le 3$, respectively.  The $(\mu,t)$ coverage increases for larger $\ell_1$, and we have full coverage for all the panels in the lower row. Taking all panels together, we have $\Delta_\sigma \le 2$ for the majority of bins. There are a few bins where $2 < \Delta_\sigma \le 3$, and only three bins where $3 < \Delta_\sigma \le 5$. In general, we may interpret the deviations between $B(\ell_1,\mu,t)$ and $[B_{\rm Ana}]_d(\ell_1,\mu,t)$ to be consistent with statistical fluctuations. 
 The three bins where $3 < \Delta_\sigma \le 5$ all occur near the squeezed limit $(\ell_1 \approx \ell_2, \ell_3 \rightarrow 0)$  where, possibly, the deviations are not entirely due to statistical fluctuations. Here, a part of the deviation may arise because the bispectrum is very sensitive to the exact triangle configuration near the squeezed limit that considers $\ell_3 \rightarrow 0$, and the value of the bispectrum changes very rapidly even within the bin \citep{shaw21,gill_2024}.

\section{Summary and Conclusion}
\label{sec:sum}
There is considerable motivation to quantify the three-point statistics of the radio sky. In this work, we have considered radio-interferometric observations, for which we present a visibility-based estimator for the angular bispectrum. 
The three-visibility correlation directly probes the bispectrum \citep{Bharadwaj2005}. However, the computational cost, which scales as the cube of the number of visibilities, makes it impractical to implement a direct correlation. Here we deal with the gridded visibilities instead. Although this reduces the computation, it still scales as the fourth power of the total number of grid points $N_t$, which can be computationally expensive. Here, we have implemented an FFT based fast estimator \citep{Sefusatti_thesis,Jeong_thesis,sco2015} where the computation time scales as $\propto N_t^2\log(N_t^2)$. Here, we follow \citet{shaw21} to present a binned angular bispectrum estimator $\hat{B}(\ell_1,\mu,t)$, where the angular multipole $\ell_1$ and the dimensionless parameters $(\mu,t)$ respectively quantify the size and shape of the triangle.  For the analysis presented in this work, it takes $\sim 5$ seconds to perform all the FFTs and $\sim$ 1 minute 10 seconds to compute the ABS corresponding to all possible triangles for a single realization on a single core CPU.

We have used the simulated $154.25 \, {\rm MHz}$ MWA observations
\citep{Patwa_2021} to validate our estimator. We have simulated visibility data considering a sky signal that has a known input model angular bispectrum. We find that these observations can be used to probe the angular bispectrum over a wide range of triangle sizes $(46 \le \ell_1 \le 1320)$, and shapes. The estimated values are found to be in good agreement with the model predictions, and the deviations between these two are largely consistent with those expected from statistical fluctuations. Our analysis validates the estimator and demonstrates that the MWA observations considered here have the potential to quantify the angular bispectrum with $\approx 10-15 \%$  accuracy. The analysis presented here does not take into account real observed data which includes foregrounds, system noise and possible systematics.
In future work, we plan to apply our estimator to analyze the actual MWA data. We also plan to generalize the estimator so as to quantify the three-dimensional bispectrum of redshifted  21-cm brightness temperature fluctuations. A proper foreground removal or avoidance and mitigation of possible systematics are crucial to detect the 21-cm  bispectrum using radio-interferometric observations.

\section*{Acknowledgements}
The authors thank the anonymous reviewer for valuable suggestions and comments. SSG acknowledges the support of the Prime Minister
Research Fellowship (PMRF).

\section*{Data Availability}

The simulated data and package involved in this work
will be shared on reasonable request to the authors.

\bibliography{main}{}
\bibliographystyle{aasjournal}

\appendix

\section{Two and three visibility Correlations}
\label{app:2vis_derivation}
We present a brief derivation of Eq.~\ref{eq:2vis} and Eq.~\ref{eq:S3}.   It is convenient to work in the continuum limit where  Eq.~\ref{eq:TbFT}, \ref{eq:maps1} and \ref{eq:mabs1} are respectively given by  
\begin{equation}
 \delta T_b(\bfth) =\int \frac{d^2 \ell}{(2 \pi)^2}  \exp[ -i \bfl \cdot \, \bfth]~\Delta \tilde{T}_b(\bfl)
~ \, , 
\label{eq:Tl_cont}
\end{equation}
\begin{equation}
 \langle~ \Delta \Tilde{T}_b({\bfl}) \, \Delta  \Tilde{T}^*_b({\bfl^\prime})~\rangle  =
 (2\pi)^{2} \, \delta^2_D(\bfl - \bfl^\prime) \, C_{{\ell}}
\label{eq:maps_cont}
\end{equation}
and 
\begin{equation}
\langle~ \Delta \Tilde{T}_b(\bfl_1) \Delta\Tilde{T}_b(\bfl_2) \Delta \Tilde{T}_b(\bfl_3)~\rangle = (2\pi)^{2}~\delta^2_D(\bfl_1 + \bfl_2 + \bfl_3)~B (\ell_1, \ell_2, \ell_3) ~\,
 \label{eq:mabs_cont}
\end{equation}
where $ \delta^2_D(\bfl - \bfl^\prime)$ is the 2D Dirac delta function. We also define the 
aperture power pattern  
\begin{equation}
    \Tilde{a}(\u) =\int \, d^2 \theta\, {A}(\bfth)  ~\exp[{i2\pi\u\cdot\bfth}] 
    \label{eq:apt}  
\end{equation}
which is the Fourier transform of  $ {A}(\bfth)$.  Using the convolution theorem, 
the visibility $\V(\u)$ (Eq.~\ref{eq:vis}) can be expressed as
\begin{equation}
\V(\u)= Q\, \int \,  \frac{d^2 \ell}{(2 \pi)^2}  \Tilde{a}(\u - {\bfl}/(2 \pi)) ~\Delta \Tilde{T}_b(\bfl) ~,
\label{eq:viscon}
\end{equation}

For the Gaussian beam ${A_G}(\bfth)$ given in Eq.~\ref{eq:A_G},  we have the Gaussian aperture power pattern
\begin{equation}\label{eq:a_tilde}
    {\Tilde{a}_G}(\u) =  \pi\theta_0^2 \, \exp[{-\pi^2 \,\theta_0^2 \,U^2}]~.
\end{equation}
The correlation of two visibilities corresponding to baselines $\u$ and $\u+\Delta\u$ is given by,
\begin{equation}
    \langle  \V(\u){\V}^*(\u+\Delta\u) \rangle \, = Q^2\, \int \,  \frac{d^2 \ell^{\prime}}{(2 \pi)^2} 
     \Tilde{a}_G(\u- {\bfl^{\prime}}/(2 \pi)) \Tilde{a}_G^*( {\u}+\Delta\u- {\bfl^{\prime}}/(2 \pi) )  C_{\ell^{\prime}}
\label{eq:a2_1}
\end{equation}
We assume that the $C_{\ell^{\prime}}$ does not change significantly within $\theta_0^{-1}$ the width of  $\Tilde{a}_G(\u- {\bfl^{\prime}}/(2 \pi))$ whereby we can hold its value constant at $\bfl^{\prime}= 2 \pi \u$ and  take it outside the integral. We then have 
\begin{equation}
    \langle  \V(\u){\V}^*(\u+\Delta\u) \rangle \, = Q^2\, \left[ \int \,   \frac{d^2 \ell^{\prime}}{(2 \pi)^2} 
     \Tilde{a}_G(\u- {\bfl^{\prime}}/(2 \pi)) \Tilde{a}_G^*( {\u}+\Delta\u- {\bfl^{\prime}}/(2 \pi) )  \right ] C_{\ell=2 \pi U}
\label{eq:a2}
\end{equation}
The expression in the square brackets is a standard two-dimensional Gaussian integral, yielding
\begin{equation}
\langle  \V(\u){\V}^*(\u+\Delta\u) \rangle \, =  \dfrac{\pi\theta_0^2 Q^2}{2} \exp   
\left [ {-\pi^2\,\theta_0^2\, \Delta U^2/2} \right] \, C_{\ell}\,  ~.
\label{eq:a3}
\end{equation}
where $\ell=2 \pi U$.

We next consider the three-visibility correlation. Using Eq.~\ref{eq:mabs_cont},  the correlation of three visibilities corresponding to baselines $\u_1,\u_2 ~\text{and}~ \u_3+\Delta\u$ is given by
\begin{equation}
    \begin{aligned}
        \left\langle \V(\u_1) \V(\u_2) \V(\u_3 + \Delta \u) \right\rangle
&= Q^3 \int \frac{d^2 \ell_1^{\prime}}{(2 \pi)^2} \, \int \frac{d^2 \ell_2^{\prime}}{(2 \pi)^2} 
 \, \tilde{a}_G(\u_1- {\bfl_1^{\prime}}/(2 \pi)) \,  
 \tilde{a}_G(\u_2- \bfl_2^{\prime}/(2 \pi))  \, \Tilde{a}_G^*( \u_3+\Delta\u - \bfl_3^{\prime}/(2 \pi) )
 \\&  \times B(\ell_1^{\prime}, \ell_2^{\prime}, \ell_3^{\prime})
    \end{aligned}
\label{eq:3vis_corr_der}
\end{equation}
where $\bfl_3^{\prime} =-\bfl_1^{\prime} - \bfl_2^{\prime}$.

Here we assume $\u_1 + \u_2  + \u_3=0$, that is, they form a closed triangle, and $\Delta\u$ quantifies the deviation from the closed triangle configuration. As in Eq.~\ref{eq:a2}, here too we assume that $B(\ell_1^{\prime}, \ell_2^{\prime}, \ell_3^{\prime})$ does not vary significantly over the width of  $\Tilde{a}_G(\u- {\bfl_1^{\prime}}/(2 \pi))$ and  $\Tilde{a}_G(\u- {\bfl_2^{\prime}}/(2 \pi))$.  This allows us to replace 
 $B(\ell_1^{\prime}, \ell_2^{\prime}, \ell_3^{\prime})$  with  $B(\ell_1, \ell_2, \ell_3)$
where $\bfl_1=2 \pi \u_1$, $\bfl_2= 2 \pi \u_2$ and $\bfl_3=2\pi\u_3$, and write it outside the integral in Eq.~\ref{eq:3vis_corr_der}. We then have 
\begin{equation}
    \begin{aligned}
        \left\langle \V(\u_1) \V(\u_2) \V(\u_3 + \Delta \u) \right\rangle
&= Q^3 \left[ \int \frac{d^2 \ell_1^{\prime}}{(2 \pi)^2} \, \int \frac{d^2 \ell_2^{\prime}}{(2 \pi)^2} 
 \, \tilde{a}_G(\u_1- {\bfl_1^{\prime}}/(2 \pi)) \,  
 \tilde{a}_G(\u_2- \bfl_2^{\prime}/(2 \pi))  \, \Tilde{a}_G^*( \u_3+\Delta\u - \bfl_3^{\prime}/(2 \pi) ) \right]
 \\&  \times B(\ell_1, \ell_2, \ell_3)
    \end{aligned}
\label{eq:3vis_corr_der_1}
\end{equation}

The terms in the square bracket are a four-dimensional Gaussian integral. We evaluate this to obtain 
\begin{equation}
\langle \V(\u_1) \V(\u_2) \V(\u_3+\Delta \u)\rangle =  \dfrac{\pi\theta_0^2 Q^3}{3} \exp \left [ {-\pi^2\theta_0^2\Delta U^2/3} \right] \,B({\ell_1}, {\ell_2}, {\ell_3}) \,.
\end{equation}
which is the relation between the three-visibility correlation and the ABS.

\section{Angular Bispectrum for our  non-Gaussian Model}
\label{app:bispectrum_derivation}
We provide a brief derivation of Eq.~\ref{eq:bsana}. The non-Gaussian model of brightness temperature fluctuations (Eq.~\ref{eq:mb1}) is given by the convolution in the Fourier space as,
\begin{equation}
\Delta \Tilde{T}_{\rm b}(\bfl) =   \Delta \Tilde{T}_{\rm G}(\bfl) + \dfrac{f_{\rm NG}}{\sigma_T}\, \frac{1}{(2\pi)^2} \int d^2 \ell_1 
 \Delta \Tilde{T}_{\rm G}(\bfl - \bfl_1)\, \Delta \Tilde{T}_{\rm G}(\bfl_1)   
\label{eq:mb2}
\end{equation}

We proceed by substituting $\Delta \Tilde{T}_{\rm b}(\bfl)$ (Eq.~\ref{eq:mb2}) into the definition of the ABS (Eq.~\ref{eq:mabs_cont}). Note that the expectation value of the product of an odd number of Gaussian random fields is zero. The leading order nonzero term, in powers of $f_{\rm NG} (\ll 1)$, is 
\begin{equation}\label{eq:3_temp_cor}
   (2\pi)^{2}  \delta^2_D(\bfl_1 + \bfl_2 + \bfl_3)~ B_{\rm Ana}(\ell_1,\ell_2, \ell_3)  =  \dfrac{f_{\rm NG}}{\sigma_T}\, \Bigg[\langle\Delta \tilde{T}_{\rm G}(\bfl_1)\Delta \tilde{T}_{\rm G}(\bfl_2)\int 
   \frac{d^2\ell'}{(2\pi)^2}
   \Delta \tilde{T}_{\rm G}(\bfl_3-\bfl')\Delta \tilde{T}_{\rm G}(\bfl')\rangle ~+~\text{2 terms} \Bigg] \,.
\end{equation}
The two other terms indicated above can be respectively obtained by interchanging $\bfl_1$ with $\bfl_3$, and $\bfl_2$ with $\bfl_3$. There are higher order terms in $f_{\rm NG} $ that we ignore.

Wick's theorem states that,
\begin{equation}\label{eq:wicks}
\begin{aligned}
    \langle\Delta \tilde{T}_{\rm G}(\bfl_1)\Delta \tilde{T}_{\rm G}(\bfl_2)\Delta \tilde{T}_{\rm G}(\bfl_3)\Delta \tilde{T}_{\rm G}(\bfl_4)\rangle &= \langle\Delta \tilde{T}_{\rm G}(\bfl_1)\Delta \tilde{T}_{\rm G}(\bfl_2)\rangle~\langle\Delta \tilde{T}_{\rm G}(\bfl_3)\Delta \tilde{T}_{\rm G}(\bfl_4)\rangle ~+
    \langle\Delta \tilde{T}_{\rm G}(\bfl_1)\Delta \tilde{T}_{\rm G}(\bfl_3)\rangle~\langle\Delta \tilde{T}_{\rm G}(\bfl_2)\Delta \tilde{T}_{\rm G}(\bfl_4)\rangle \\&+
    \langle\Delta \tilde{T}_{\rm G}(\bfl_1)\Delta \tilde{T}_{\rm G}(\bfl_4)\rangle~\langle\Delta \tilde{T}_{\rm G}(\bfl_2)\Delta \tilde{T}_{\rm G}(\bfl_3)\rangle 
    \end{aligned}
\end{equation}
Using Eqs.~\ref{eq:wicks} and \ref{eq:maps_cont} in \ref{eq:3_temp_cor}, we have,

\begin{equation}
    \begin{aligned}
    (2\pi)^{2}~& \delta^2_D(\bfl_1 + \bfl_2 + \bfl_3)~     B_{\rm Ana}(\ell_1,\ell_2, \ell_3)   =\dfrac{f_{\rm NG}}{\sigma_T}\, \Bigg[ \int  
    d^2\ell^{\prime} \,  {(2\pi)^2}\, \big\{ \delta^{2}_D\,(\bfl_1 + \bfl_2) ~\delta^{2}_D\,( \bfl_3) ~ C_{\ell_1} \,   C_{\ell^\prime}\,\\&
    +\delta^{2}_D\,(\bfl_1 + \bfl_3 - \bfl') ~\delta^{2}_D\,(\bfl_2 +  \bfl') ~ C_{\ell_1} \,   C_{\ell_2}\,+
    \delta^{2}_D\,(\bfl_1 + \bfl') ~\delta^{2}_D\,(\bfl_2 + \bfl_3 - \bfl') ~ C_{\ell_1} \,   C_{\ell_2}\,
    \big\}  ~+~\text{2 terms} \Bigg]. 
    \end{aligned}
\end{equation}

The first term on the RHS  only contributes when $\bfl_3=0$. This refers to the $\bfth$ 
independent constant component of  $\delta T_b(\bfth)$, which is zero and can be ignored. 
Performing the $\bfl'$ integral and including the other two terms, we obtain the final expression for the ABS to the first order in $f_{NG}$,
\begin{equation}
B_{\rm Ana}(\ell_1,\ell_2, \ell_3) = \dfrac{2\,f_{\rm NG}}{\sigma_T} \Big({C_{\ell_1}}\,{C_{\ell_2}}+{C_{\ell_2}}\,{C_{\ell_3}}+ {C_{\ell_3}}\,{C_{\ell_1}}\Big) \,,
\end{equation}
where the three $\bfl$ modes form a closed triangle, i.e., $\bfl_1 + \bfl_2 + \bfl_3=0$.

\label{lastpage}
\end{document}